\documentclass[twocolumn]{aastex63}
\usepackage{lineno}
\usepackage{hyperref}
\usepackage[utf8]{inputenc}
\usepackage[reqno]{amsmath}
\usepackage{amsfonts}
\usepackage{enumerate}
\usepackage{ytableau}
\usepackage{seqsplit}
\usepackage{multirow}
\usepackage{mathtools}
\usepackage{subfigure}
\usepackage{graphicx}
\usepackage{indentfirst}
\usepackage[english]{babel}
\usepackage{overpic}
\usepackage[normal,footnotesize]{caption2}
\def\baselinestretch{1.25}
\usepackage{appendix}
\usepackage{amssymb}
\usepackage{amsthm}
\usepackage{epsfig,graphicx,epstopdf}
\usepackage{natbib}
\usepackage{scalerel}

\def\deg{\hbox{$^\circ$}}
\newcommand{\fermi}{\textit{Fermi}-{\rm LAT}}
\newcommand{\planck}{\textit{Planck}}

\newcommand\HI{H{\footnotesize I}}
\newcommand\HII{H{\footnotesize II}}

\received{***}
\revised{***}
\accepted{***}
\submitjournal{ApJL}

\shorttitle{cosmic ray variation in Danks 1 and 2}
\shortauthors{Liu et.al}

\begin{document}
\title{Detection of extended gamma-ray emission in the vicinity of Cl Danks 1 and 2}
   \author{Jia-hao Liu}
   \affiliation{CAS Key Laboratory for Research in Galaxies and Cosmology, Department of Astronomy, School of Physical Sciences, University of Science and Technology of China, Hefei, Anhui 230026, China}
   \affiliation{School of Astronomy and Space Science, University of Science and Technology of China, Hefei, Anhui 230026, China}
   \author{Bing Liu}
   \affiliation{CAS Key Laboratory for Research in Galaxies and Cosmology, Department of Astronomy, School of Physical Sciences, University of Science and Technology of China, Hefei, Anhui 230026, China}
   \affiliation{School of Astronomy and Space Science, University of Science and Technology of China, Hefei, Anhui 230026, China}
   \author{Rui-zhi Yang}
   \affiliation{CAS Key Laboratory for Research in Galaxies and Cosmology, Department of Astronomy, School of Physical Sciences, University of Science and Technology of China, Hefei, Anhui 230026, China}
   \affiliation{School of Astronomy and Space Science, University of Science and Technology of China, Hefei, Anhui 230026, China}
   \affiliation{Tianfu Cosmic Ray Research Center, Chengdu, Sichuan 610213, China}
   \correspondingauthor{Rui-zhi Yang}
   \email{yangrz@ustc.edu.cn}

   \date{Received XXX; accepted YYY; in original form \today}
\begin{abstract}
We report the detection of high-energy gamma-ray emission towards the G305 star-forming region. Using almost 15 years of observation data from {\sl Fermi} Large Area Telescope, we detected an extended gamma-ray source in this region with a significance of $\sim 13 \sigma$.  The gamma-ray radiation reveals a clear pion-bump feature and can be fitted with the power law parent proton spectrum with an index of $-2.5$. The total cosmic ray (CR) proton energy in the gamma-ray production region is estimated to be the order of $10^{49}\ \rm erg$. We further derived the CR radial distribution from both the gamma-ray emission and gas distribution and found it roughly obeys the $1/r$ type profile, which is consistent with other similar systems and expected from the continuous injection of CRs by the central powerful young massive star cluster Danks 1 or Danks 2 in this region.  Together with former detections of similar gamma-ray structures, such as Cygnus cocoon, Westerlund~1, Westerlund~2, NGC~3603, and W40, the detection supports the hypothesis that young massive star clusters are  CR accelerators.
\end{abstract}


\label{firstpage}



\section{Introduction}
\label{sec:intro}

The origin of cosmic rays (CRs) remains an open question in astrophysics. The recent advances in gamma-ray astronomy indicate that the young massive star clusters (YMCs) can be an alternative type CR sources \citep{2019Nature} in addition to supernova remnants (SNRs).  Furthermore, several independent observations also hint at such a hypothesis. For example, recent findings by \cite{2016binns} suggest that a significant portion of CRs may be accelerated in young OB star clusters and associated super bubbles, as indicated by measurements of ${ }^{60} \mathrm{Fe}$ abundance in CRs. Additionally, studies on the Galactic diffuse gamma-ray emission reveal that the radial distribution of CRs aligns more closely with OB stars rather than SNRs, as reported by \cite{2016acero} and \cite{2016Yang}. Superbubbles, fueled by supernova explosions and collective stellar winds, are capable of providing the necessary kinetic energy to generate the observed flux of CRs locally, as proposed by \cite{2004parizot}. These structures are expected to emit gamma rays due to the interaction of freshly accelerated CRs with the surrounding gas. In recent days, YMCs have become popular candidates for CR accelerators in several studies. The Cygnus cocoon is a typical example of such sources, despite the Cygnus OB system not being the most potent young star cluster in our Galaxy. \cite{cygnus1} have studied the \fermi data of the Cygnus cocoon and provided an illustration for investigating the early stages of CR youth within a superbubble setting before they integrate into the older Galactic population. Other YMC examples include Westerlund 2\citep{2018yang}, W40\citep{2020sun}, W43\citep{w43}, Carina\citep{carina}, Rosette\citep{rosette}, G25\citep{RSGC1}, and NGC 3603\citep{2017yang}. Characteristics of CRs in such regions are well studied and described, making YMCs important sites for studying CRs.

The G305 complex (G305.4+0.1)  is identified as one of the most massive star-forming structures in the Galaxy. It is located in the Scutum-Crux arm ($l=305.4^\circ$, $b=+0.1^\circ$; \citealt{coordinate}) at an estimated distance of $\sim$4 kpc, and was delineated by both mid- and far-infrared, submillimetre and radio emission \citep{danks12}. The presence of methanol masers and ultracompact  \HII\ regions (G305~\HII\ complex) indicates ongoing massive star formation in the G305 complex. It has the form of a large trilobed cavity with a maximum extent of $\sim$34 pc, delineated by both mid- and far-infrared, submillimetre, and radio emission. The existence of embedded massive young stellar objects and a substantial reservoir of cold molecular gas available for future activity is observed in the G305 complex. Similar properties are also observed in the star-forming complex associated with NGC 3603 and 30 Dor.

The young clusters Danks 1 and 2 are situated at the core of the star-forming complex. The age of Danks 1 is $1.5_{-0.5}^{+1.5} \mathrm{Myr}$, and the age of Danks 2 is $3_{-1}^{+3} \mathrm{Myr}$ \citep{danks12}. The complex's overall morphology strongly suggests several epochs of sequential star formation, which have been initiated and sustained by the influence of these two central clusters \citep{danks12}. The G305 complex's morphology strongly suggests multiple epochs of sequential star formation, a process initiated and sustained by the influence of the two central clusters. Photometric investigations of these clusters were undertaken by \cite{2004Bica} and \citep{2009Baume}, albeit somewhat impeded by significant visual extinction and densely populated fields. Regarding the distance to Danks 1 and Danks 2, we assume they are located at a distance of 4kpc, the same distance as the G305 complex \citep{danks12}.

In this paper, we analyzed the 15 years of \fermi\ data in this region and found hard gamma-ray emission associated with radius from the center YMCs Danks 1 and Danks 2. We also derived the radial distribution of CR energy density, which is consistent with the former results. This paper is organized as follows. In Sec.\ref{sec:gas}, we studied the distribution of three phases of gas content around Danks 1 and Danks 2. In Sec.\ref{sec:fermi}, we analyzed \fermi\  data and studied the spatial and spectral distribution of gamma-ray emissions. In Sec.\ref{sec:CR}, we derived the CR spectrum utilizing analysis results of Sec.\ref{sec:fermi} and Sec.\ref{sec:gas}, and investigated the distribution of CR energy density over radius. Finally, in Sec.\ref{sec:dis}, we discussed our results along with former works and drew our conclusions.

\section{Gas content around Danks 1 and Danks 2}
\label{sec:gas}

We investigated three different gas phases, i.e., the H$_{2}$, the neutral atomic hydrogen (\HI), and the \HII, in the vicinity of Danks 1 and Danks 2. We followed the method in Sec.3 of \cite{carina} to derive the gas column density map of these three gas phases scaled to H column density.

First, we used the CO composite survey data \citep{Dame01}  to trace the $\rm H_{2}$. We took $N({\rm H_{2}}) = X_{\rm CO} \times W_{\rm CO}$ \citep{Lebrun1983}, where 
$X_{\rm CO}$ is the $\rm H_{2} / CO$ conversion factor that chosen to be $\rm 2.0 \times 10^{20}\ cm^{-2}\ K^{-1}\ km^{-1}\ s$ as suggested by \cite{Dame01} and \cite{Bolatto13}. 

Then for the \HI\ density, we used the data-cube of the \HI\ $\rm{4\pi}$ survey   (HI4PI), which is a 21-cm all-sky database of Galactic \HI\ \citep{HI4PI16}. 
We estimated the \HI\ column density using the equation,

\begin{equation}
N_{\rm HI} = -1.83 \times 10^{18}T_{\rm s}\int \mathrm{d}v\ {\rm ln} \left(1-\frac{T_{\rm B}}{T_{\rm s}-T_{\rm bg}}\right),
\end{equation}

where $T_{\rm bg} \approx 2.66\ \rm K$ is the brightness temperature of the cosmic microwave background radiation at 21 cm, and $T_{\rm B}$ is the brightness temperature of the \HI\ emission. 
In addition, in the case when $T_{\rm B} > T_{\rm s} - 5\ \rm K$, we truncate $T_{\rm B}$ to $T_{\rm s} - 5\ \rm K$, where $T_{\rm s}$ is chosen to be 150 K. 

\begin{figure*}
        \centering
        \includegraphics[width=0.3\linewidth]{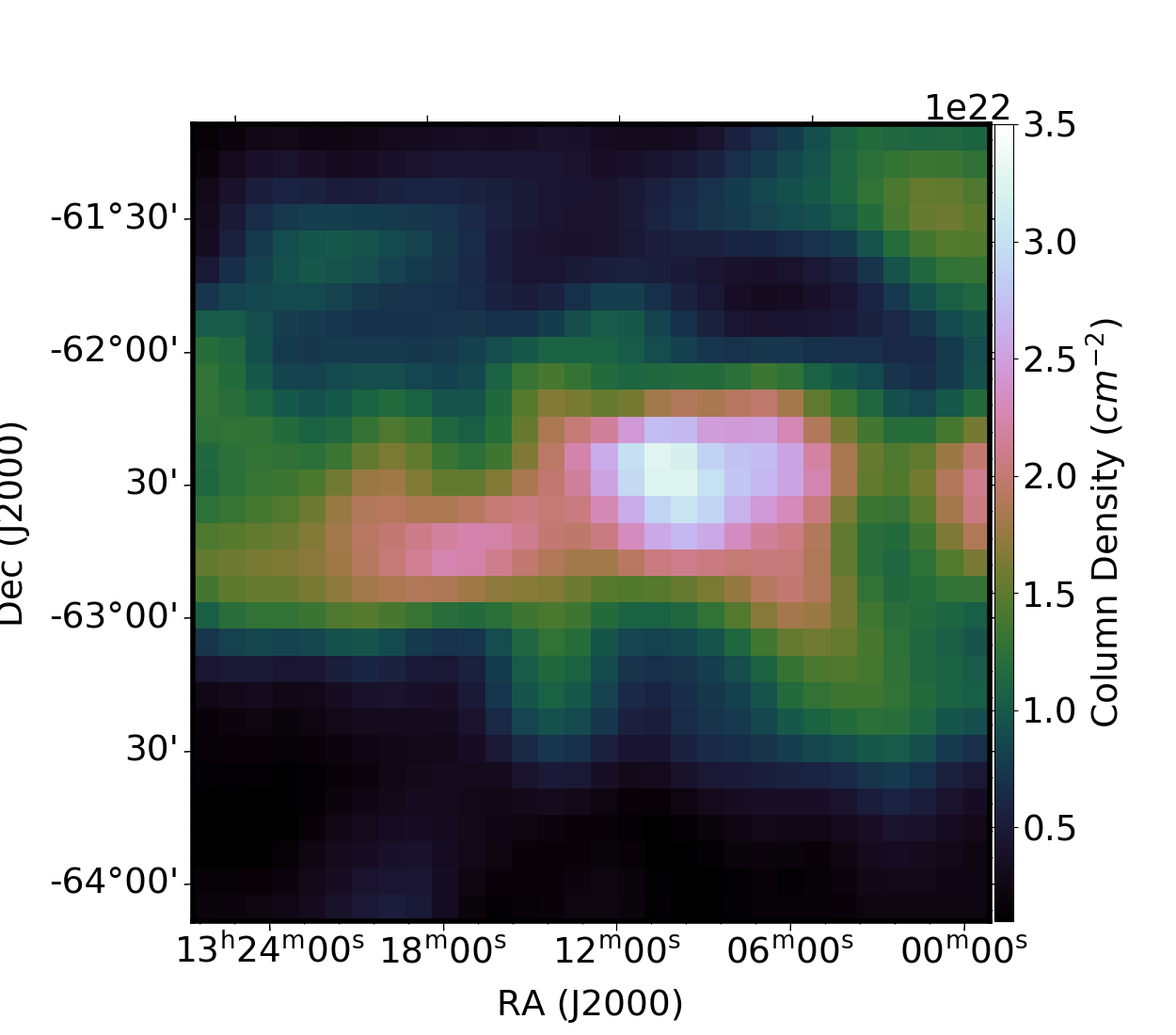}
        \includegraphics[width=0.3\linewidth]{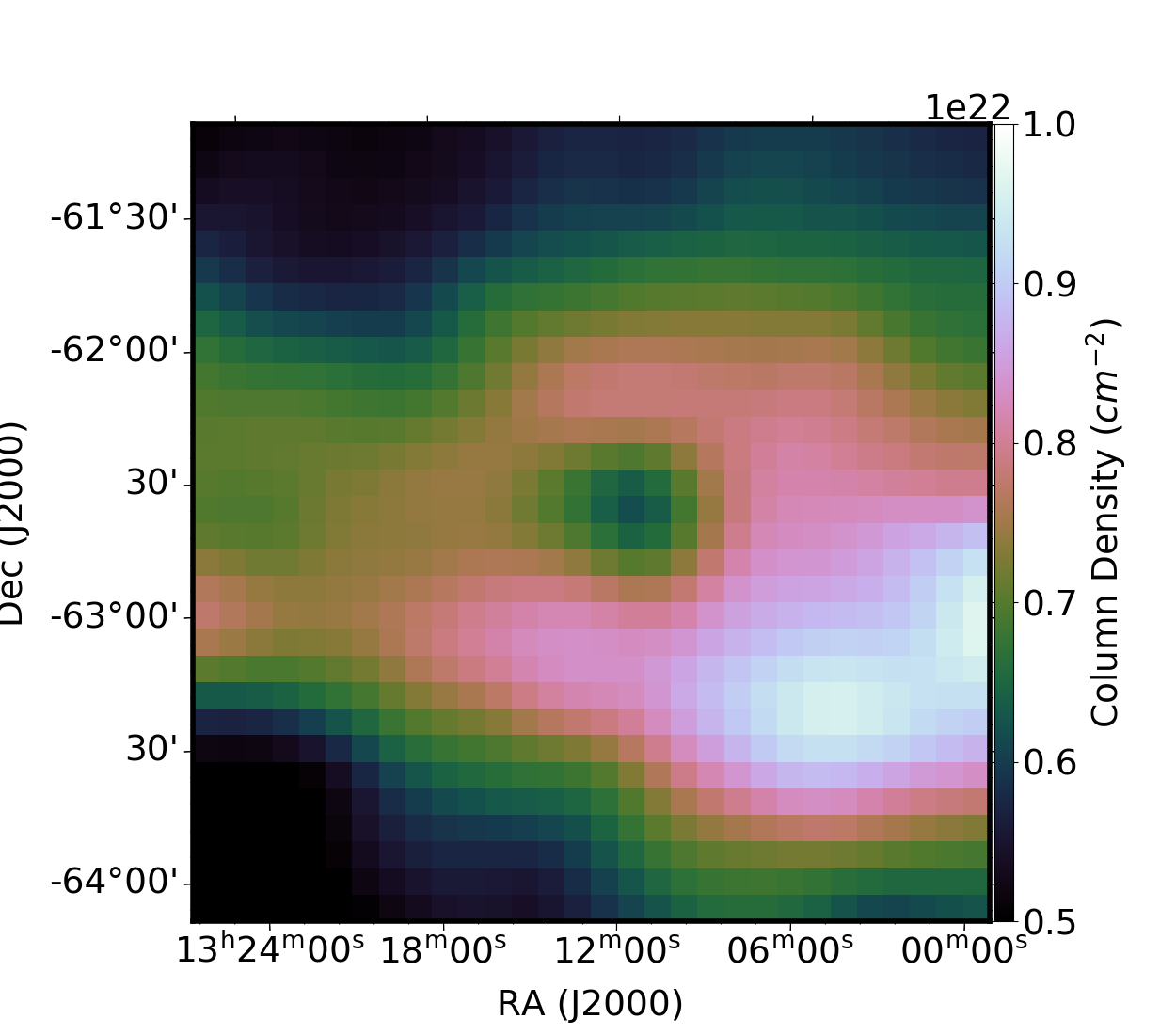}
        \includegraphics[width=0.3\linewidth]{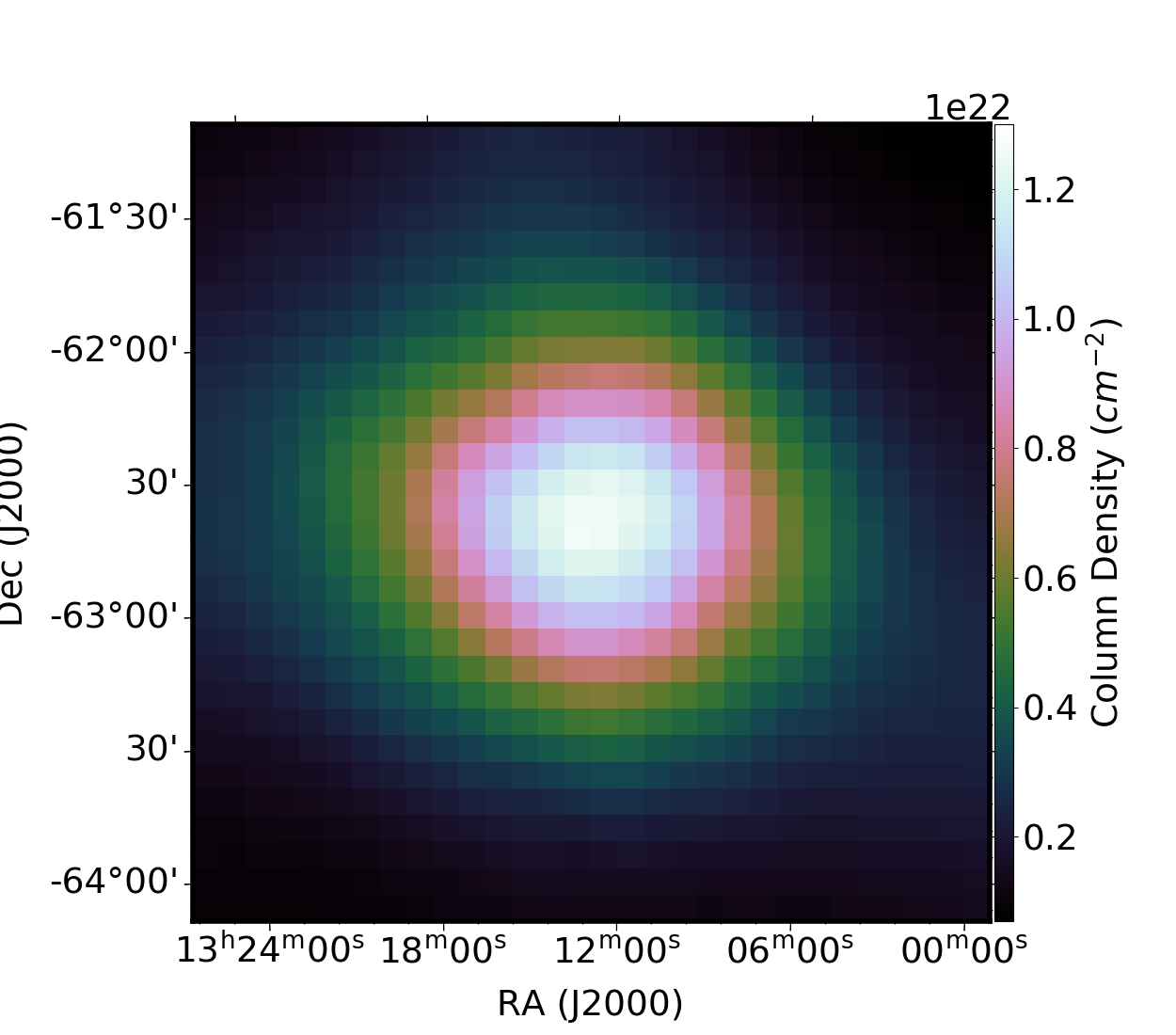}
    \caption{Maps of gas column densities in three gas phases. Left shows the H$_{2}$ column density derived from the CO data \citep{Dame01}. Middle shows the \HI\ column density derived from the HI4PI survey. Right shows the \HII\ column density derived from the \planck\ free-free map assuming the effective density of electrons $n_{\rm e}=10~\rm cm^{-3}$.} \label{fig:gas} 
\end{figure*}

The radial velocity ranges of Danks 1 and Danks 2 was estimated to be $[-35.37, 6.01]$~km~s$^{-1}$ and $[-46.35, 10.51]$~km~s$^{-1}$ by \cite{v_dis}, respectively. In this study, we took the velocity range [-46.35, 10.51]~km~s$^{-1}$ for integration of the H$_2$ and \HI\ data. The total gas mass and mean column density of the H atom in these two phases are shown in Tab.\ref{tab:gas}. The integrated maps of the H atom column density in these two phases are shown in the left and middle panels of Fig.\ref{fig:gas}.

G305 region is one of the largest ionized gas complexes in our Galaxy, thus the ionized gas can also make a significant contribution to the total gas mass.  For the \HII\ column density, we used the \planck\ free-free map \citep{Planck16}. 
First, we transformed the emission measure into free-free intensity using the conversion factor in Table 1 of \cite{Finkbeiner03}. 
Then, we calculate the \HII\ column density from the intensity ($I_{\nu}$) of free-free emission by using Eq.(5) of \cite{Sodroski97}, 
\begin{equation}
\begin{aligned}
  N_{\rm HII} = &1.2 \times 10^{15}\ {\rm cm^{-2}} \left(\frac{T_{\rm e}}{1\ \rm K}\right)^{0.35} \left(\frac{\nu}{1\ \rm GHz}\right)^{0.1}\left(\frac{n_{\rm e}}{1\ \rm cm^{-3}}\right)^{-1} \\
&\times \frac{I_{\nu}}{1\ \rm Jy\ sr^{-1}},
\end{aligned}
\end{equation}
in which the frequency $\nu = \rm 353\ GHz$  and the electron temperature $T_{e} =\rm 8000\ K$. 
The \HII\ column density is inversely proportional to the effective density of electrons $n_{\rm e}$. 
Thus, we adopted an effective density $10\ \rm cm^{-3}$, which is the value suggested in \cite{Sodroski97} for the region inside the solar circle. The total gas mass and mean column density of the H atom in \HII\ phase are shown in Tab.\ref{tab:gas}. The map of H atom column density in \HII\ phase is shown in right panel of Fig.\ref{fig:gas}.

\begin{table}
\caption{Estimated gas mass and column density for three gas phases.}  
\label{tab:gas}       
\centering                        
\begin{tabular}{ccc}       
\hline\hline     
phase&mass($M_\odot$)&mean column density(cm$^{-2}$)\\
\hline  
H$_{2}$&1.33$\times10^6$&1.64$\times10^{22}$\\
\HI&6.24$\times10^5$&7.68$\times10^{21}$\\
\HII&6.56$\times10^5$&8.09$\times10^{21}$\\
\hline
total&2.61$\times10^6$&3.22$\times10^{22}$\\
\hline
\end{tabular}
\end{table}

\section{\fermi\ data analysis}
\label{sec:fermi}
To study the gamma-ray emission in the vicinity of Danks 1 and 2, we used the latest \fermi\ Pass 8 data from August 4, 2008 (MET 239557417) until May 22, 2023 (MET 706481557), and used the Fermitools from Conda distribution\footnote{\url{https://github.com/fermi-lat/Fermitools-conda/}} for the data analysis. 
In this work, we chose a 20\deg $\times$ 20\deg\ square region centered at the position of Danks 1 and 2 (R.A. = 198.222$\deg$, Dec. = -62.681$\deg$) as the region of interest (ROI).
We utilized {\sl gtselect} to choose "source" class events (evtype = 3 and evclass = 128) with zenith angles below 90$^{\circ}$ to eliminate background contamination from the Earth's limb. The good time intervals were determined by applying the expression $\rm (DATA\_QUAL > 0) \&\& (LAT\_CONFIG == 1)$. The analysis of the SOURCE events was conducted using the instrument response functions {\it P8R3\_SOURCE\_V3}. 
\begin{figure*}
    \centering
    \includegraphics[width=0.3\linewidth]{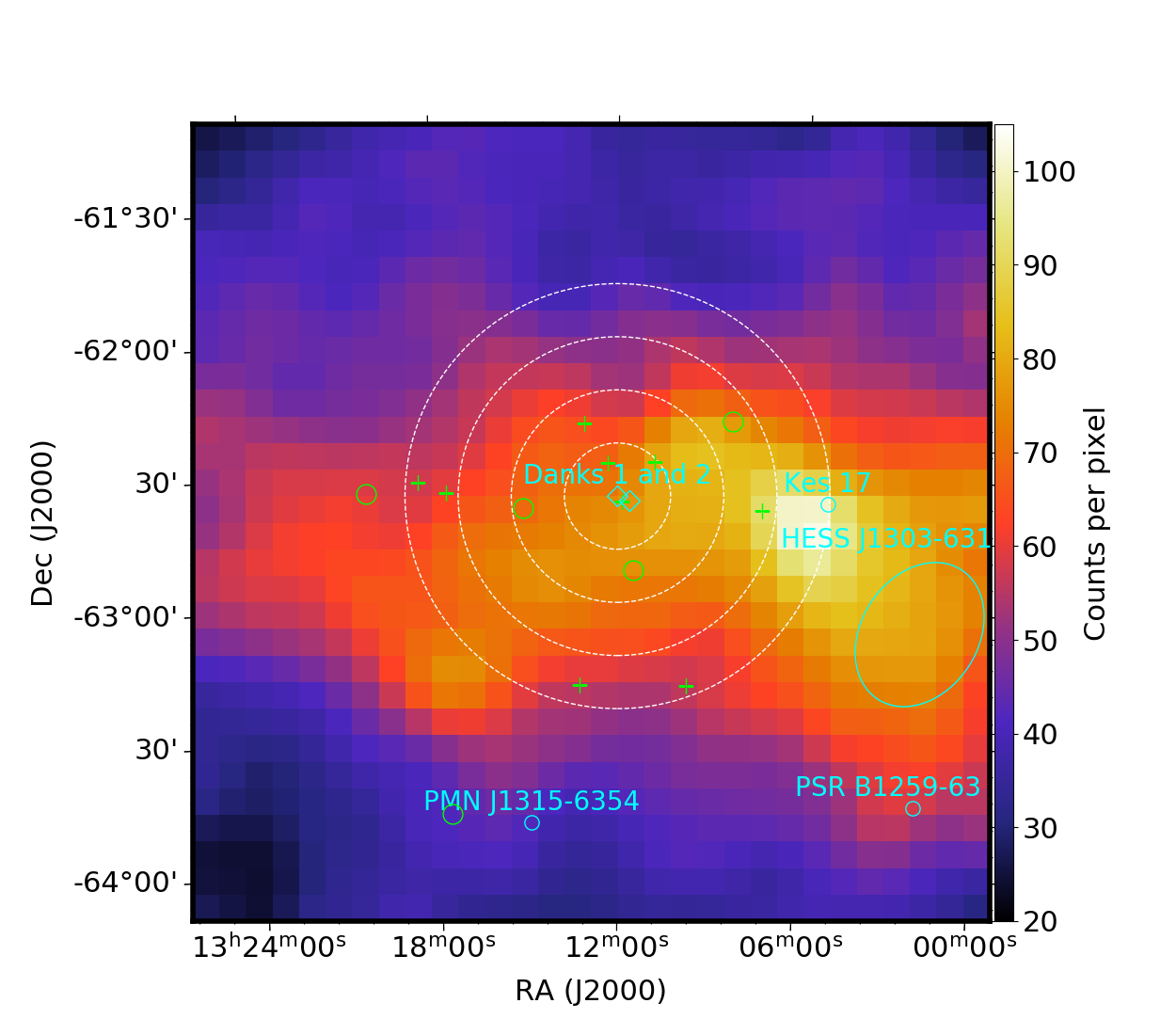}
    \includegraphics[width=0.3\linewidth]{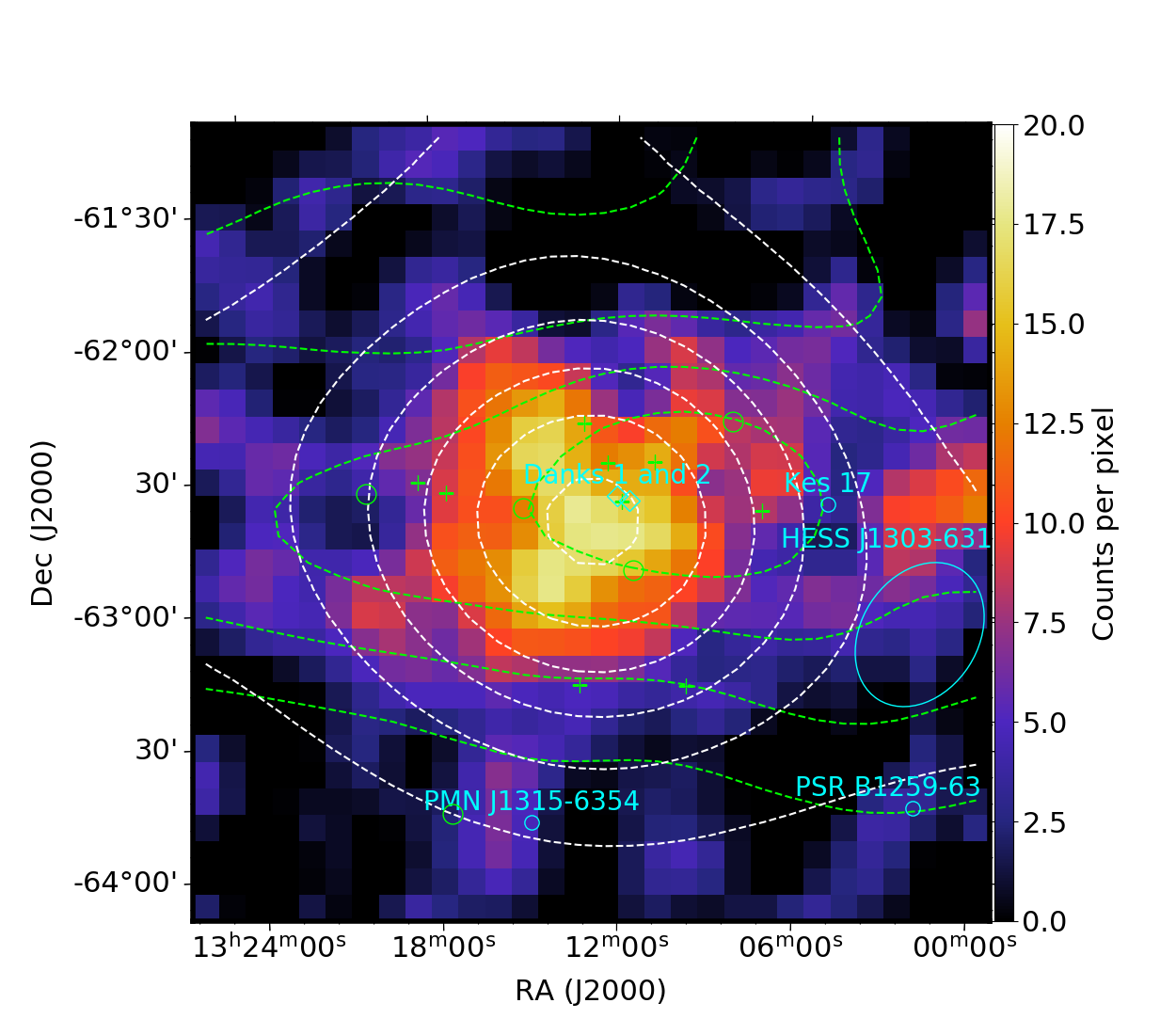}
    \includegraphics[width=0.3\linewidth]{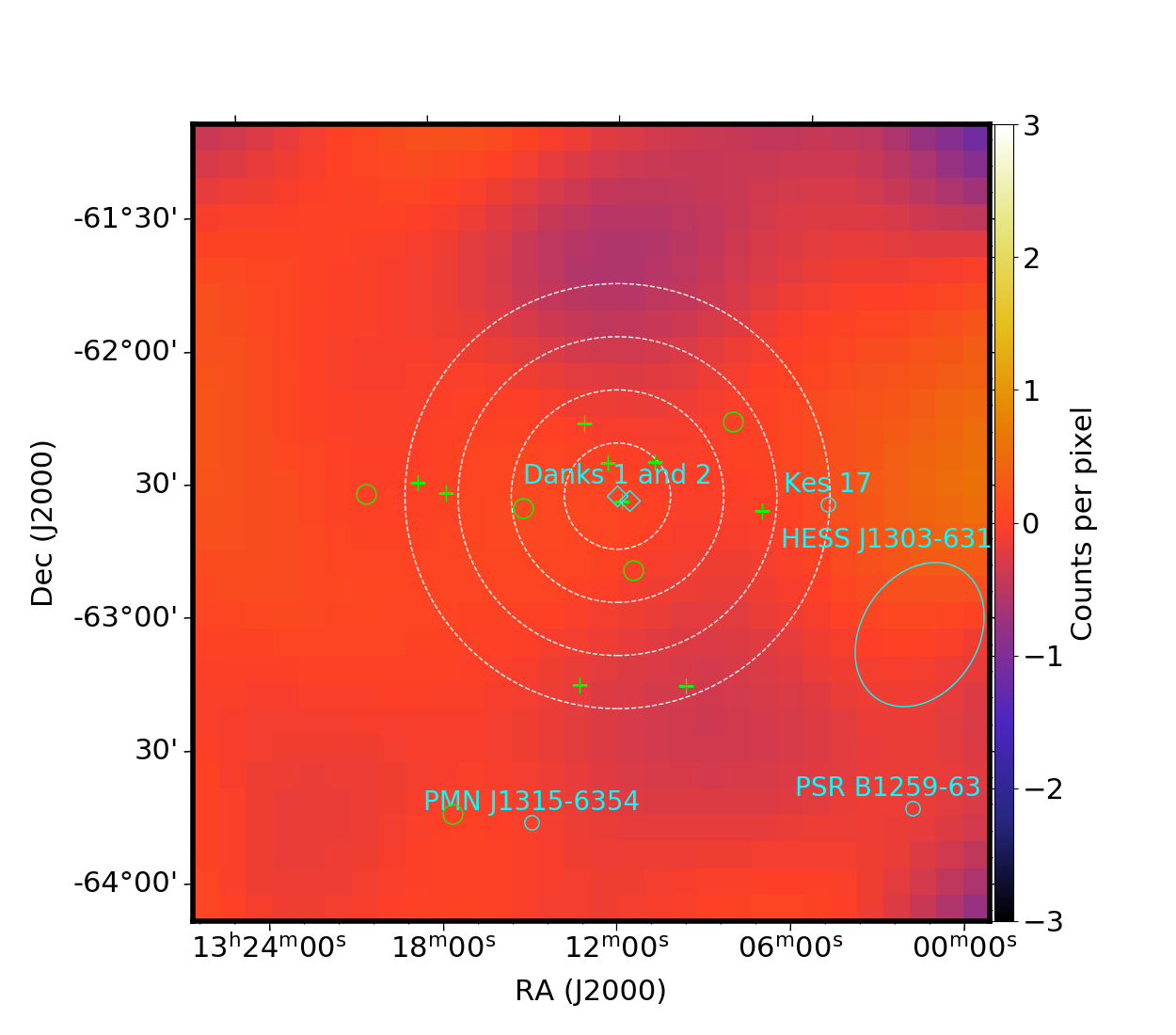}
    \caption{The left panel displays the gamma-ray counts map in the vicinity of Danks 1 and 2. The cyan diamonds indicate the positions of Danks 1 and 2. The cyan ellipse and circles represent the locations of 4FGL catalog sources with associations, while the green circles denote the positions of 4FGL sources without associations. Additionally, the green crosses display the positions of clusters located within 0.8$^\circ$ from the ROI center. The white circles illustrate the 1 disk and 3 rings discussed in Sec.\ref{sec:CR_r}. The middle panel showcases the gamma-ray residual counts map after subtracting the contribution of the diffuse background and sources other than the Gaussian disk. In the right panel, the residual significance maps from 0.1 to 500 GeV of the radial Gaussian disk model are presented, smoothed with a Gaussian filter of 0.3$^{\circ}$. }\label{fig:maps} 
\end{figure*}

A standard binned analysis was carried out following the official tutorial of binned likelihood analysis\footnote{\url{https://fermi.gsfc.nasa.gov/ssc/data/analysis/scitools/binned_likelihood_tutorial.html}}. The source models were generated using make4FGLxml.py\footnote{\url{https://fermi.gsfc.nasa.gov/ssc/data/analysis/user/make4FGLxml.py}}, which included sources from the LAT 12-year Source Catalog (4FGL-DR3, \cite{4FGL}) within the ROI enlarged by 10$^{\circ}$, the Galactic diffuse background emission (gll\_iem\_v07.fits), and the isotropic emission background (iso\_P8R3\_SOURCE\_V3\_v1.txt). The spectral parameters of sources within 10$\deg$ from the center and the normalization factor of the diffuse and isotropic background were set free.

In our ROI, there are several sources in the 4FGL catalog. Their positions are shown in Fig.\ref{fig:maps}. During the likelihood analysis, we deleted the sources without associations in the source model to avoid underestimating the flux of Danks 1 and Danks 2.
\subsection{Spatial analysis}
\label{sec:spatial}
We first used the events from 1 to 500~GeV to study the spatial distribution of the gamma-ray emission in the ROI. The gamma-ray counts map in the $3\deg \times 3\deg$ region of the vicinity of Danks 1 and 2 is shown in the left panel of Fig.\ref{fig:maps}. The clusters mentioned in Sec.\ref{sec:intro} are marked in the counts map. There are three sources in 4FGL-DR3 in the ROI with corresponding identified sources (4FGL J1305.5-6241 and Kes 17, 4FGL J1315.9-6349c and PMN J1315-6354, HESS J1303-631), which are shown in blue marks. Other sources in 4FGL-DR3 are shown in green crosses.

With the sources in the 4FGL-DR3 catalog, we conducted a binned likelihood analysis and applied the fitted 4FGL-DR3 model to calculate the residual map in the ROI. 
Evident excess around the center clusters was found from the residual map, indicating the presence of gamma-ray emission excess in the region. We then added a radial Gaussian template centered at the position of Danks 1 and 2 (R.A. = 198.222$\deg$, Dec. = -62.681$\deg$) to account for the emission near Danks 1 and 2. We calculated the log-likelihood -$\log({\cal L})$ assuming different sigma of the Gaussian disk and found that the best fit is achieved when sigma equals 0.5$^{\circ}$ (-$\log({\cal L})$=21538). The significance of this added source was calculated to be $\sim$13$\sigma$. We also tested another spacial template generated using the total gas column density map derived in Sec.\ref{sec:gas}. To build this template, we normalized the three different gas phases, i.e., the H$_{2}$, the neutral atomic hydrogen (\HI), and the \HII, to the column density of H atom($N_H$), and summed them together. However, the likelihood fitting result using this gas template (-$\log({\cal L})$=21620) is worse than the Gaussian disk. Furthermore, we utilized the Akaike Information Criterion (AIC) to determine which models best suited the data. The AIC is computed by subtracting twice the logarithm of the likelihood from twice the number of free parameters (k) in the model, expressed as  $-2 \log({\cal L})+2 \mathrm{k}$. The comparison of fitting results of different models can be seen in Tab.\ref{tab:like}. So in the later analysis, we used the radial Gaussian disk as the spacial template for  the gamma-ray excess around Danks 1 and 2. We generated the model maps with this source model using {\sl gtmodel}, and obtained the residual counts map by subtracting the contribution of the diffuse background and sources other than the Gaussian disk, which is shown in the middle panel of Fig.\ref{fig:maps}. We also derived the total residuals for the ROI by subtracting the model maps from the observed gamma-ray counts maps. We then divided the residual counts by the square root of the observed gamma-ray counts, i.e., the residual significance ((signal-to-noise, S/N, in $\sigma$), of each pixel. The result is shown in the right panel of Fig.\ref{fig:maps}.
\begin{table}
\caption{Fitting result of different source models}  
\label{tab:like}       
\centering                        
\begin{tabular}{cccc}       
\hline\hline     
model&-$\log({\cal L})$&free parameters&AIC\\
\hline  
4FGL&21620&79&43398\\
+ Gaussian disk&21538&81&43238\\
+ gas template&21543&81&43248\\
\hline
\end{tabular}
\end{table}

\subsection{spectral analysis}
\label{sec:spectral}
\begin{figure}  
    \centering
        \includegraphics[width=\linewidth]{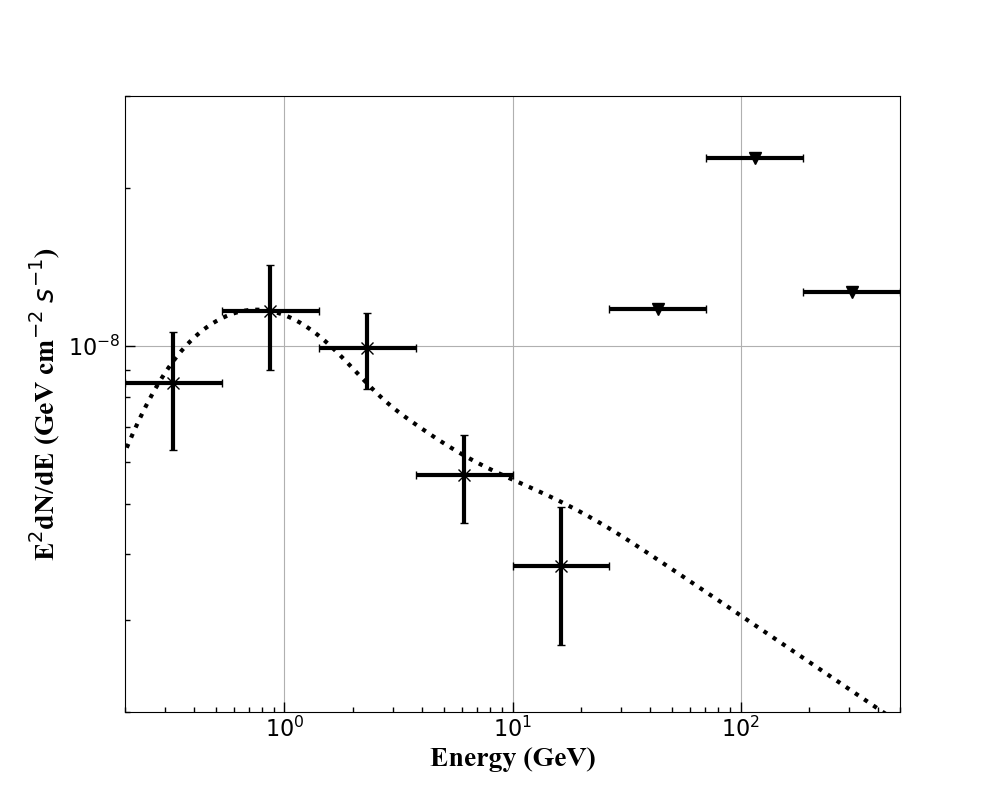}
    \caption{SED of gamma-rays from the region near Danks 1 and 2. The error bar of each data point indicates the uncertainty caused by the statistics and the systematic error. The inverted triangles indicate the 3-sigma upper limits. The dashed lines show the corresponding gamma-ray emission from the pion-decay mechanism assuming the best-fit CR spectra (see Section~\ref{sec:CR}).} \label{fig:gamma}
\end{figure}

To study the gamma-ray spectrum from the vicinity of Danks 1 and 2, we divided the data from 200 MeV to 500 GeV into 8 energy bins evenly distributed in logarithmic space. For the data in each energy bin, we conducted likelihood analysis with the source model in Sec.\ref{sec:spatial} and derived the total flux of the region near Danks 1 and 2. Systematic uncertainties of the flux for each energy bin were also estimated using the method described on the Fermi official website \footnote{\url{https://fermi.gsfc.nasa.gov/ssc/data/analysis/scitools/Aeff_Systematics.html}}. The derived gamma-ray spectrum is shown in Fig.\ref{fig:gamma}. The total gamma-ray luminosity was estimated to be $\sim1.0\times10^{35}~\rm erg~s^{-1}$. Besides,  we calculated the 95\% flux upper limit for energy bins with the source test statistic (TS) value less than 4, those are shown in inverted triangles. We note that the energy flux of the first energy bin shows a clear hint of a low energy break, which is consistent with the pion-bump feature in the hadronic scenario of gamma-ray productions.

\section{CR content in the vicinity of Danks 1 and 2}
\label{sec:CR}
Due to the clear pion-bump feature, We assumed that the gamma-ray emission from Danks 1 and 2 is mainly generated from the proton-proton interaction between CR protons and the gases. Under such assumptions, we studied the CR content around Danks 1 and 2 utilizing gamma-ray data from Sec.\ref{sec:fermi} and gas distributions derived from Sec.\ref{sec:gas}.
\subsection{CR spectral analysis}
We assumed the CR protons follow the momentum distribution function: $F=f_ip^{\Gamma}$, where $F$ is the flux density of protons in the unit of $\mathrm{cm}^{-2}\, \mathrm{s}^{-1}\,\mathrm{GeV}^{-1}$,  $p$ is the proton momentum in the unit of GeV/c, and $\Gamma$ stands for the spectral index. 

Under this assumption, we used the gamma-ray production cross section from \citet{2014pp} to perform a likelihood fitting to the gamma-ray data. We also calculated the gas mass in the gas template we used in Sec.\ref{sec:spatial} so that we can derive the absolute CR fluxes. We found that the best-fit CR proton spectral index $\Gamma$ is 2.36$\pm$0.09. 
The corresponding gamma-ray spectrum is shown in the dashed line in Fig.\ref{fig:gamma}. 
\subsection{The radial distribution of CR protons}
\label{sec:CR_r}
Using both the gas and gamma-ray spatial distributions, we can derive the spatial distribution of CRs, which contain unique information on the injection and propagation of CRs.   If the CRs are injected continuously from the central source as in some other YMC systems\citep{2019Nature}, then in regions near the source, the energy distribution of CR protons/electrons $F(R) \sim \frac{1}{R D}$\citep{2022yang}, in which R is the radial distance and D is the diffusion coefficient of CRs. Consequently, CR energy density $\Omega_\mathrm{CR}(R)$ would also be proportional to 1/R, i.e.,
\begin{equation}
\Omega_\mathrm{CR}(R) = \omega_0\left(\frac{R}{R_0}\right)^{-1},
\label{eqcr}
\end{equation}
where $R_0$ is the size of the emission region taken to be 56~pc($\sim$0.8\deg), and $\omega_0$ is the CR energy density at distance $R_0$.
Due to the projection effect, the 2D radial distribution function of CR energy density can be calculated as follows: 
\begin{equation}
\omega_\mathrm{CR}(r)=2 \int_r^{R_0} \frac{\Omega(R) R}{\sqrt{R^2-r^2}} d R,
\end{equation}
in which $r$ is the projected radial distance. 
We can then derive:
\begin{equation}
\omega_\mathrm{CR}(r) =\omega_0 \log \left(\left(R_0+\sqrt{R_0^2-r^2}\right) / r\right) / \sqrt{R_0^2-r^2}.
\end{equation}

\begin{figure}
    \centering
        \centering
        \includegraphics[width=0.9\linewidth]{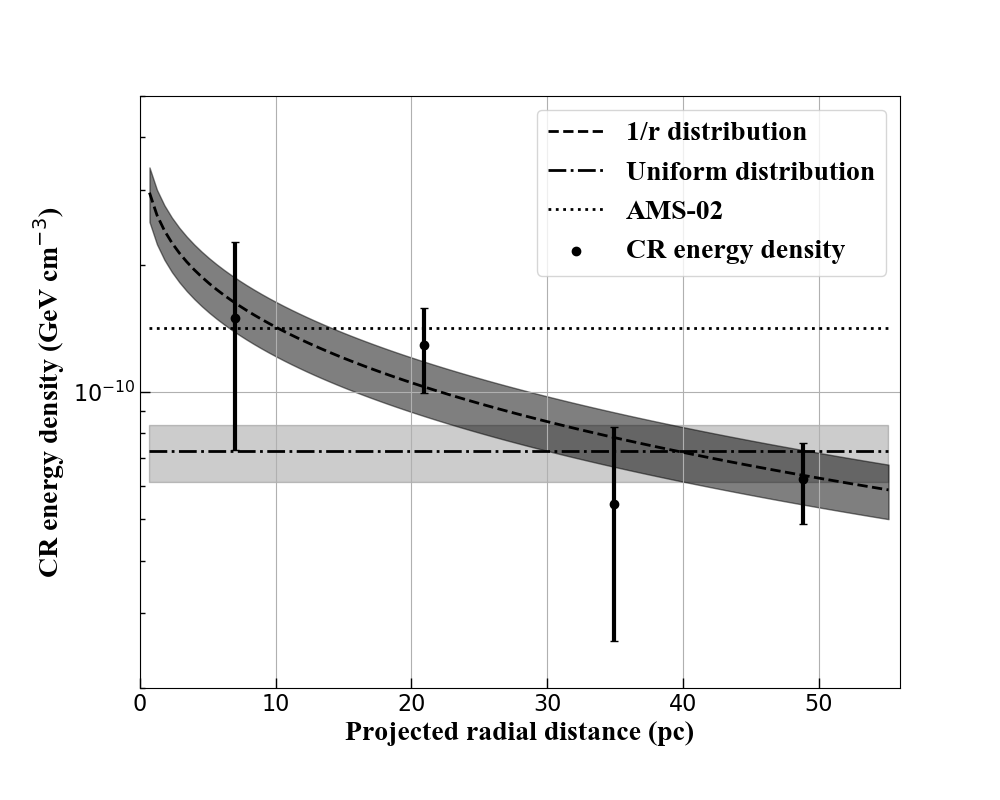}    
    \caption{Radial distribution CR energy density over 10 GeV in the region near Danks 1 and 2. The shaded area indicates the uncertainty caused by the statistics and the systematic error. The dotted line indicates the local CR proton energy density over 10 GeV measured by AMS-02 \citep{Aguilar15}.} \label{fig:cr_r}
\end{figure}

To test whether the CR energy density in the vicinity of Danks 1 and 2 follows the distribution described above, we divided the emissive region into four regions: a central disk with a radius of 14~pc(approximately 0.2$^\circ$), and three concentric rings extending from 14~pc to 28~pc (approximately 0.2$^\circ$ to 0.4$^\circ$), 28~pc to 42~pc (approximately 0.4$^\circ$ to 0.6$^\circ$), and 42~pc to 56~pc (approximately 0.6$^\circ$ to 0.8$^\circ$), respectively. We replaced the radial Gaussian disk with these four flat spacial templates and applied the same spatial analysis with that of 
Sec.\ref{sec:spatial} to derive the gamma-ray flux of each region. Next, assuming the spectral shape of CRs is uniform in the whole region, we calculated the CR energy density of each region.  Applying maximum likelihood estimation, we then derived that $\omega_0=0.1~\rm eV~cm^{-3}$. To prove that this is a better assumption than an assumption of a uniform distribution of CR, 
we calculated $\chi^2$ of each distribution and obtained $\chi^2/d.o.f=5.66/3$ for the uniform profile and $\chi^2/d.o.f=1.52/3$ for the $1/r$ profile. From the statistics, the $1/R$ distribution is favored marginally. 
The comparison of the fitting results of two assumptions is shown in Fig.\ref{fig:cr_r}. We also calculated the local CR energy density by integrating the local CR proton flux measured by AMS-02 \citep{Aguilar15}, which is also shown in Fig.\ref{fig:cr_r} as a comparison. In addition, we integrated Eq.\ref{eqcr} and derived the total CR energy $E_\mathrm{CR}$:
\begin{equation}
    E_\mathrm{CR}=2\pi \omega_0R_0 R_{\rm dif}^2=1.7\times10^{49}~ (\frac{R_{\rm dif}}{100~\rm pc})^2\mathrm{erg}, 
\end{equation}
where $R_{\rm dif}$ is the diffusion length of the CRs injected by the central accelerator. It should be noted that $R_{\rm dif}$ can be significantly larger than $R_{0}$, which represents the radius of the gamma-ray emission region. Since the size of the measured gamma-ray emission is limited by both the gas distribution and the gamma-ray instrument sensitivity, the CRs can indeed occupy a much larger volume. Assuming the age of Danks 1 of $T\sim 1.5$ Myr and a diffusion coefficient of $D \sim 10^{28}~\rm cm^2/s$, $R_{\rm dif}$ can be estimated as $R_{\rm dif} \sim 2 \sqrt{DT} \sim 400~\rm pc$, which is much larger than the gamma-ray emission region detected in GeV band. 

\section{Discussion and conclusion}
\label{sec:dis}
In this paper,  by analyzing almost 15 years of \fermi data, we found extended gamma-ray emission in the vicinity of Danks 1 and 2, which can be modeled with a radial Gaussian disk. 
Then we derived the SED of the extended gamma-ray emission and found a bump feature, which is a strong hint that the gamma-ray emissions are produced via the pion-decay process in the interaction of CR protons with ambient gas. A power-law distribution of CR protons with an index of $\sim2.5$ can typically reproduce the gamma-ray SED. 

We found that the Gaussian model is superior to the gas distribution model while modeling the gamma-ray emission, which may indicate the distribution of gamma-ray emission is primarily influenced by the radial distribution of CRs rather than the spatial distribution of gas. We obtained the gamma-ray fluxes at various radii by applying the spatial templates of 1 disk and 3 rings, then calculated the variation of CR energy density across different radii. We found that the CR energy density around Danks 1 and 2 decreases with radius and its radial distribution is generally consistent with $1/r$ profile as other similar structures in \citet{2019Nature} considering the projection effect, which also implies a continuous injection of CRs as in other YMC systems. 

Given the extended emission and hard spectrum, the gamma-ray emission in this region is more likely produced by the CR injected by the YMC interacting with the ambient gas, which is similar to other gamma-ray emission YMCs\citep{2019Nature}. The YMCs Danks 1 and Danks 2 can both be the CR accelerators. Indeed, both clusters are powerful enough to accelerate the CRs needed to explain the observed gamma-ray emissions.  Danks 1 is associated with 6 WR stars, while Danks 2 is associated with 1 WR star \citep{WR_stars}. The winds of these WR stars alone can provide the kinetic power to accelerate the CRs, and dozens of OB stars in these two clusters can provide similar power.   The total kinematic power can be estimated roughly as $(3-5) \times 10^{38} \rm erg/s$. Thus, taken into account the age of about 1.5 Myr, the total kinetic energy injected is more than $10^{52} ~\rm erg$. On the other hand, the total CR energy derived in Sec.\ref{sec:CR_r} is about $10^{50}~\rm erg$ even if we assume the diffusion of CRs near this region is as fast as in the Galactic plane, which may imply an acceleration efficiency of at most $1\%$. Indeed, the diffusion coefficient near the CR sources is expected to be much smaller \citep{2019Nature}, and in this case, the acceleration efficiency can be orders of magnitude smaller. 

The wind power in Danks 1 and Danks 2 is as strong as in other powerful YMCs, such as Cygnus OB2 \citep{cygnus1,2019Nature} and NGC 3603 \citep{2017yang}, but the gamma-ray luminosity of about $10^{35} \rm erg/s$ is nearly one order of magnitude smaller, which also imply a lower CR acceleration efficiency in this region. On the other hand, this system is similar to W40 \citep{2020sun}, regarding the very young age (about 2 Myr) and high gas density ($n\sim 100~\rm cm^-{3}$). The gamma-ray emission in W40 also reveals a spectral index of about $-2.4$ \citep{2020sun}, which is similar to the results we derived here, but significantly softer than other YMCs ($2.1-2.2$) \citep{2019Nature}.  For W40 the derived total CR energy budget is of $10^{47}~\rm erg$ which also indicates a quite low acceleration efficiency than in other systems. So W40 and Danks 1/Danks 2 may represent a new sub-type of YMC systems, in which CRs are accelerated in a lower efficiency and softer spectrum. 

Future gamma-ray observations with higher angular resolution and multiwavelength study are crucial to understanding the particle acceleration and confinement in these systems. In this regard the Einstein Probe \citep{ep} which is already launched and the forthcoming CTA \citep{cta}  can provide unique information to pinning down the radiation mechanism and shed light on the mystery of the origin of CRs.

\section{Acknowledgements}

Rui-zhi Yang is supported by the NSFC under grant 12041305 and 12393854, and the national youth thousand talents program in China.
Bing Liu acknowledges the support from the NSFC under grant 12103049.

\bibliographystyle{aasjournal}
\bibliography{apjl}

\label{lastpage}
\end{document}